\documentclass[12pt,iop,preprint,tighten,apj]{aastex}

\usepackage{multirow,color}
\usepackage{bbding}
\usepackage{amssymb}
\usepackage{abstract}
\usepackage{array}
\usepackage{multicol}
\usepackage{textcmds}

\def\gsim{\;\lower4pt\hbox{${\buildrel\displaystyle >\over\sim}$}\;}
\def\lsim{\;\lower4pt\hbox{${\buildrel\displaystyle <\over\sim}$}\;}
\def\grls{\;\lower4pt\hbox{${\buildrel\displaystyle >\over <}$}\;}

\newcommand{\soho}{{\em SOHO{}}}
\newcommand{\sdo}{{\em SDO{}}}
\newcommand{\comp}{{CoMP{}{}}}
\newcommand{\aia}{{AIA{}{}}}
\newcommand{\hinode}{{\em Hinode}}

\newcommand{\alfvenic}{Alfv\'{e}nic}
\newcommand{\pref}{\protect\ref}

\begin{document}

\shorttitle{}
\shortauthors{}
\title{On the Parallel and Perpendicular Propagating Motions Visible in Polar Plumes: An Incubator For (Fast) Solar Wind Acceleration?}
\author{Jiajia Liu\altaffilmark{1,2}, Scott W. McIntosh\altaffilmark{2}, Ineke De Moortel\altaffilmark{3}, Yuming Wang\altaffilmark{1}}
\altaffiltext{1}{Earth and Space Science School, University of Science and Technology of China, NO. 96, JinZhai Road, Hefei, China}
\altaffiltext{2}{High Altitude Observatory, National Center for Atmospheric Research, P.O. Box 3000, Boulder, CO 80307, USA}
\altaffiltext{3}{School of Mathematics and Statistics, University of St Andrews, St Andrews, Fife, KY16 9SS, UK}

\maketitle

\begin{abstract}
We combine observations of the Coronal Multi-channel Polarimeter (CoMP) and the Atmospheric Imaging Assembly (AIA) onboard the {\em Solar Dynamics Observatory} (\sdo{}) to study the characteristic properties of (propagating) \alfvenic{} motions and quasi-periodic intensity disturbances in polar plumes. This unique combination of instruments highlights the physical richness of the processes taking place at the base of the (fast) solar wind. The (parallel) intensity perturbations with intensity enhancements around 1\% have an apparent speed of 120 km/s (in both the 171 \AA{} and 193 \AA{} passbands) and a periodicity of 15 minutes, while the (perpendicular) \alfvenic{} wave motions have a velocity amplitude of 0.5 km/s, a phase speed of 830 km/s, and a shorter period of 5 minutes on the {\em same} structures. These observations illustrate a scenario where the excited \alfvenic{} motions are propagating along an inhomogeneously loaded magnetic field structure such that the combination could be a potential progenitor of the magnetohydrodynamic turbulence required to accelerate the fast solar wind.
\end{abstract}


\section{Introduction}
Polar plumes are one of the most striking features in polar coronal holes \citep[e.g.,][]{Newkirk1968}. Indeed, their relatively long lifetime and high contrast compared to the background coronal hole made them a favorite target of the \soho{} era \citep[e.g.,][]{Wilhelm2000, Banerjee2009}. They are thought of as sources of dense plasma in the fast solar wind \citep[e.g.,][]{Gabriel2003} that result from the relentless magneto-convective forcing of the upper solar atmospheric plasma \citep[e.g.,][]{Wang1998}.

(Quasi-)periodic upward-propagating intensity perturbations have been observed in various regions of the Sun \citep[e.g.,][]{Lites1999, Yamauchi2003, Banerjee2009, DeMoortel2002, Liu2012} and have widely been interpreted as propagating compressional (slow) MHD waves \citep[e.g.,][]{Ofman1999, Nakariakov2006, Banerjee2009, IDM09}.  Recent progress studying spectroscopic observations (using a ``blue-wing asymmetry" technique) of these (quasi-periodic) perturbations in active and quiescent regions reveals a more complex nature - it appears that at least part of the observed intensity perturbations can be attributed to mass motions \citep[e.g.,][]{McIntosh_DePontieu_2009, DePontieu2009, Tian2011}.  Based on the similarity analysis of the periodicity, velocity and temperature between the quasi-periodic perturbations observed in solar polar plumes and those in other regions \citep[active regions, coronal holes and the quiet Sun;][]{McIntosh_DePontieu_2009, DePontieu2009}, \cite{McIntosh2010} identified them as upward mass flows, which could be part of the supply of hot plasma to the fast solar wind \citep[][]{Parker1991}. We refer the interested reader to the review by \cite{IDM2012} and the forward modelling results of \cite{IDM2015} for an extended discussion on the difficulties of distinguishing between the propagating waves and quasi-periodic upflows model.

\alfvenic{} motions in the solar atmosphere (and in polar plumes) remained undetected until the last decade when their presence was revealed in high-resolution imaging of the chromosphere with {\em Hinode}/SOT \citep[][]{DePontieu2007b} and the unique imaging spectroscopy capability of the Coronal Multi-channel Polarimeter \citep[][]{Tomczyk2007}. How these ubiquitous waves relinquish their abundant energy to the heating and/or acceleration of the plasma in the closed and open magnetic regions of the outer solar atmosphere is not well established, although a considerable body of theory (focusing on the idea of turbulence) exists \citep[e.g.,][]{Velli1993, Verdini2010, Cranmer2005}. Recently, \cite{DeMoortel2014} and \cite{Liu2014} presented observations from CoMP which indicated that excess high frequency power (compared to levels expected from theoretical models) was present in counter-propagating (low-frequency) \alfvenic{} waves near the apex of large (trans-equatorial) coronal loops. These authors proposed that the relentless counter-propagation of the waves could be a potential reservoir of energy in the coronal system through MHD turbulence and cascade of wave energy from low to high frequencies. 

In this paper, we present polar plume observations made by CoMP and the Atmospheric Imaging Assembly \citep[AIA;][]{Lemen2012} on the {\em Solar Dynamics Observatory}, exploiting their sensitivity to motions transverse to and in the plane of the sky, respectively. As in \cite{Threlfall2013}, high-speed \alfvenic{} motions and (relatively) low-speed (longitudinal) are found to co-exist on the plumes and each process has a different periodicity. These observations shed light on the richness of the physical environment at the base of  the (fast) solar wind and are used to illustrate a scenario supporting the generation of \alfvenic{} turbulence for the fast solar wind.

\begin{figure*}
\begin{center}
\includegraphics[width=0.9\hsize]{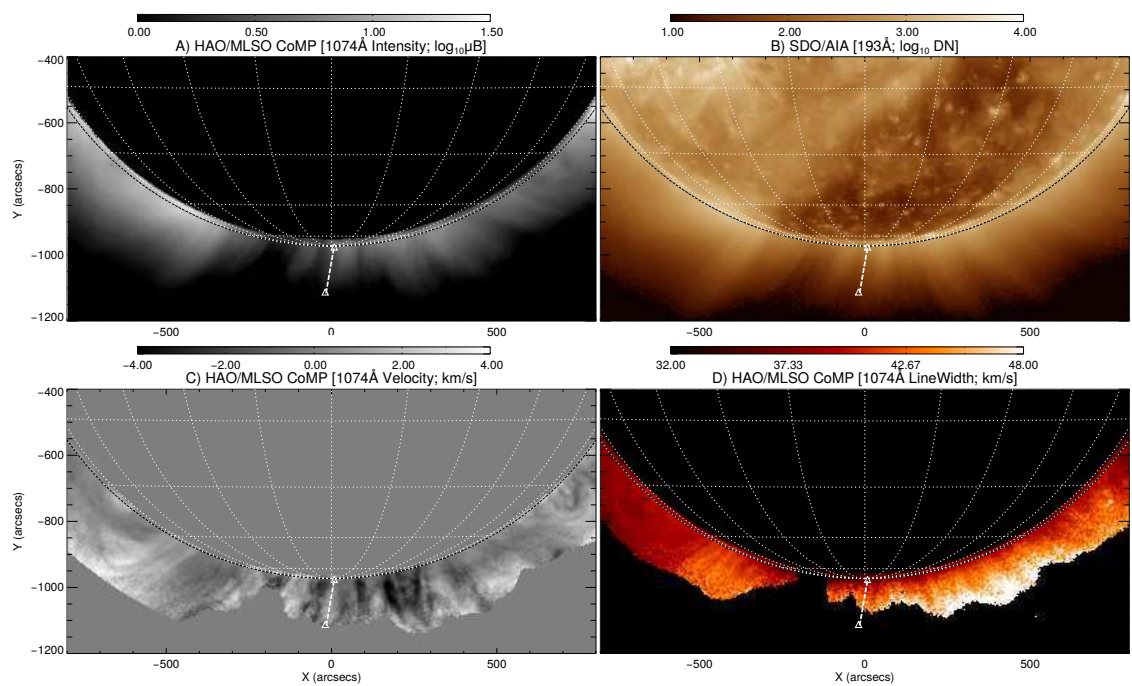}
\caption{Context images for the AIA and CoMP observations of December 30, 2011. The different panels show the CoMP \ion{Fe}{13} 10747 \AA{} peak intensity (A), Doppler velocity (C),  linewidth (D), and the AIA (\ion{Fe}{12}) 193 \AA{} intensity (B). The dashed line drawn near the south pole is the reference location for the space-time plots of Fig.~\pref{f2},~\pref{f3} and~\pref{f4}. The online edition of the Journal contains an animation of the plasma evolution over the 88 minutes of the combined observations.}\label{f1}
\end{center}
\end{figure*}

\section{Observations}\label{inst}

The Coronal Multi-channel Polarimeter \citep[CoMP;][]{Tomczyk2008} is a combination polarimeter and narrowband tunable filter that can measure the complete polarization state in the vicinity of the 10747 \AA{} and 10798 \AA{} \ion{Fe}{13} coronal emission lines. The CoMP observations used in this paper were obtained in three wavelengths (10745.2, 10746.5, and 10747.8 \AA{}) across the 10747 \AA{} \ion{Fe}{13} line, with an exposure time of 250 ms at each position. Fitted data resulted in line peak intensity, Doppler velocity, line width and enhanced intensity, all with a final cadence of 30 seconds. The images have a full field-of-view (FOV) of 2.8 R$_{\sun}$ and a spatial sampling of 4.5\arcsec{}.

AIA provides full disk images of the solar atmosphere, with high temporal cadence (12 s) and high spatial resolution (1.2\arcsec{} per pixel), extending to 1.5 solar radii. Seven narrow-band UV filters observe the Sun in a wide range of temperatures, from tens of thousands to tens of millions Kelvin, covering the atmosphere from the chromosphere to the corona. In this study, we use data from the \ion{Fe}{9} 171 \AA{} and \ion{Fe}{12} 193 \AA{} passbands, which mostly resolve material at (lower) coronal temperatures. The standard SolarSoft IDL aia$\_$prep routine is used to read and calibrate the AIA data.

\section{Analysis}\label{ana}

Figure~\pref{f1} provides context images taken on December 30, 2011 of the (south) polar plumes we investigate in detail, with the CoMP \ion{Fe}{13} 10747 \AA{} peak intensity, Doppler velocity, line width and the AIA \ion{Fe}{12} 193 \AA{} intensity in panels (A), (C), (D) and (B), respectively. A movie of the four timeseries together with that of the AIA \ion{Fe}{9} 171 \AA{} is available in the online edition of the Journal. The higher spatial resolution of the AIA observations is evident from a comparison with the CoMP intensity image.

\begin{figure}[tbh!]
\begin{center}
\includegraphics[width=0.87\hsize]{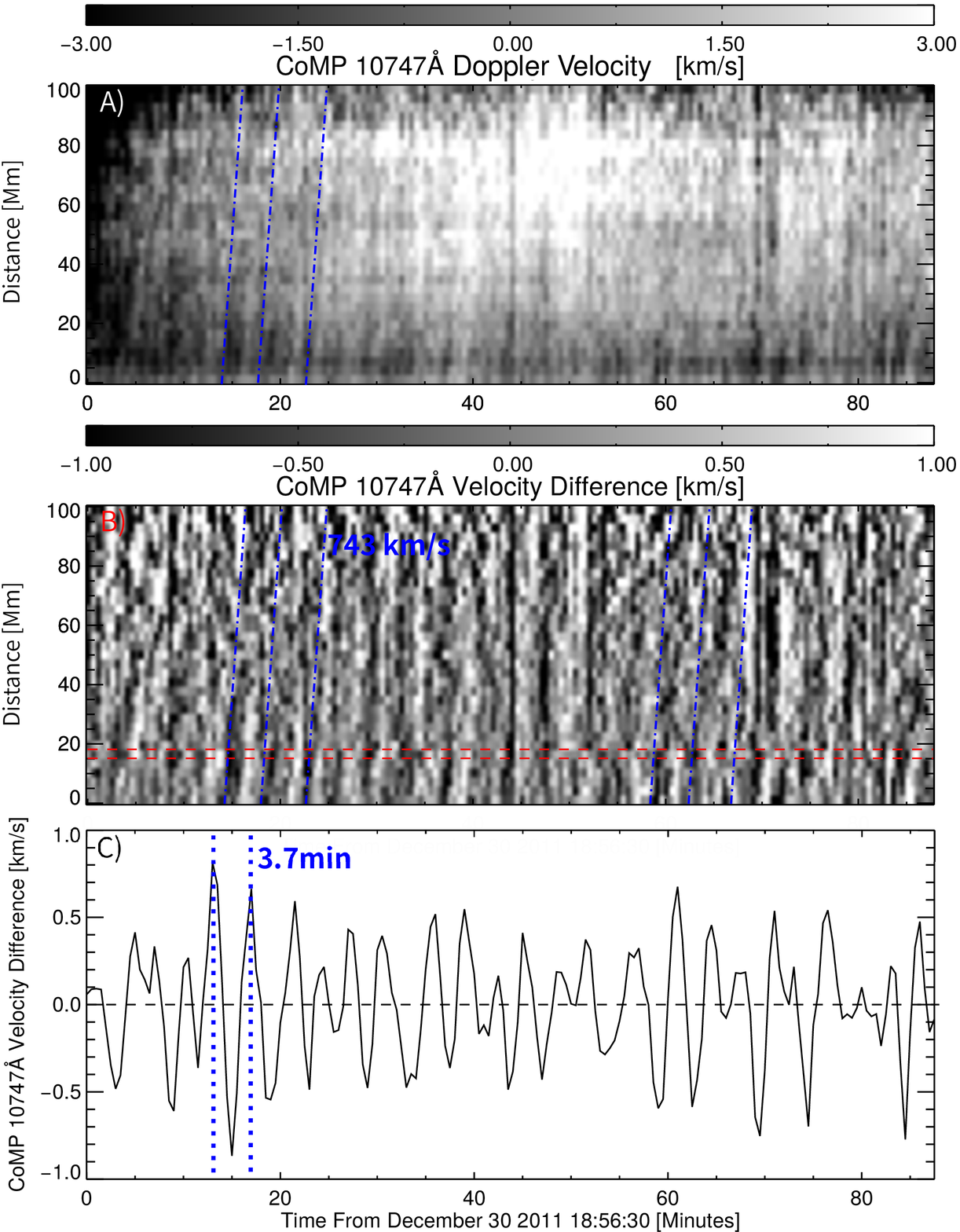}
\end{center}
\caption{Panel (A): The space-time plot from the raw data of the CoMP Doppler velocity along the dashed track highlighted in Fig.~\pref{f1}. Panel (B): Corresponding space-time plot after subtracting a 5-min-smoothed version from the original CoMP Doppler velocity data along the same track. The oblique blue dash-dotted lines in these two panels indicate several example features of propagating Doppler velocity perturbations. Panel (C): The averaged velocity difference over the region confined by the two red dashed lines in panel (B). The vertical blue dotted lines indicate a typical period of the velocity perturbations.}\label{f2}
\end{figure}

\begin{figure}[!ht]
\begin{center}
\includegraphics[width=0.9\hsize]{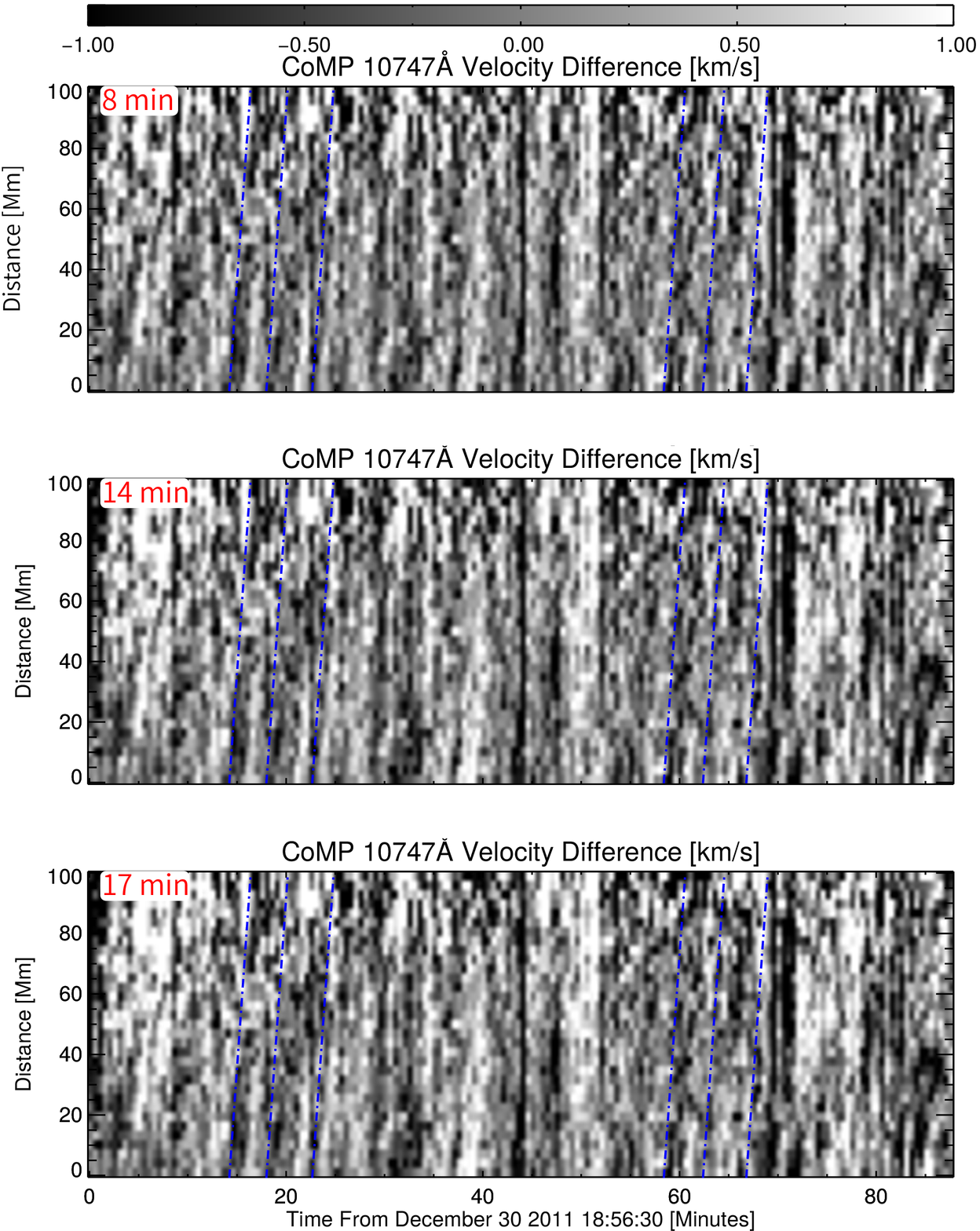}
\end{center}
\caption{Space-time plots as panel (B) in Fig.~\pref{f2} but after subtracting 8-min, 14-min and 17-min-smoothed versions, respectively. The oblique blue, dash-dotted lines are exactly the same as those in Fig.~\pref{f2} (B).}\label{f3}
\end{figure}

\begin{figure}[!ht]
\begin{center}
\includegraphics[width=0.9\hsize]{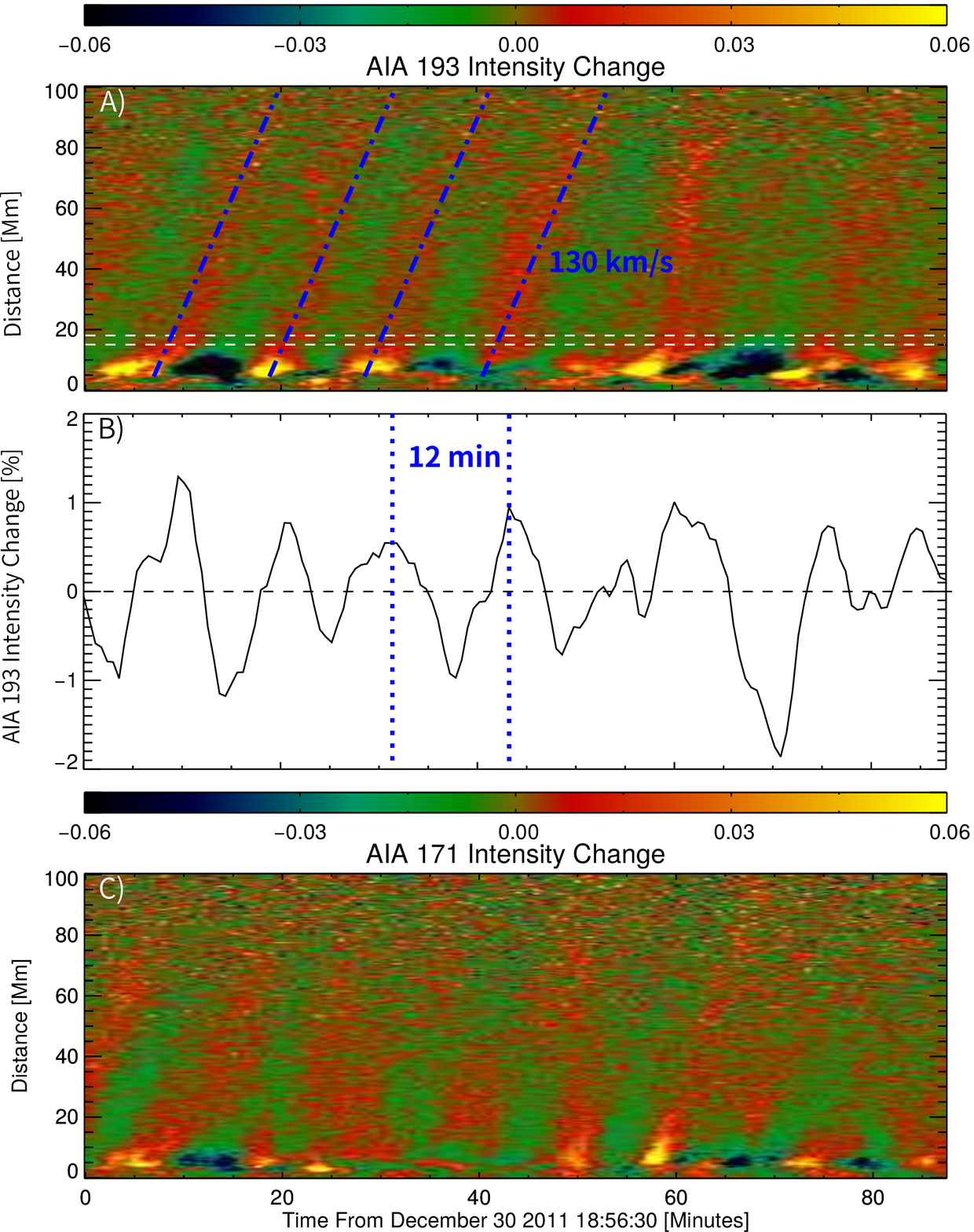}
\end{center}
\caption{Panel (A): The space-time plot of the AIA 193 \AA{} intensity changes along the dashed track highlighted in Fig.~\pref{f1}. The diagonal blue dash-dotted lines indicate the phase speeds of the intensity perturbations. Panel (B): The averaged velocity difference over the region outlined by the two white dashed lines in panel (A). The vertical blue dotted lines indicate a typical period of the intensity perturbations. Panel (C): Corresponding space-time plot of the AIA 171 \AA{} intensity changes along the same track.}\label{f4}
\end{figure}

The dashed line drawn near the south pole outlines the track for the space-time plots in Figures~\pref{f2},~\pref{f3} and~\pref{f4}. Figure~\pref{f2} (A) shows the space-time plot of the CoMP Doppler velocities along this track using the raw data. Propagating features indicating recurring Doppler velocity perturbations with periodicity from 3 to 8 minutes can be easily seen in the space-time diagram. The oblique blue dash-dotted lines indicate three examples of these perturbations. The lines have been shifted slightly to the right of the diagonal features in order not to obscure them. To further enhance the visibility of the Doppler velocity perturbations, a 10-timestep (5 minutes) smoothed version is subtracted from the original data (Doppler velocity difference, Fig.~\pref{f2} (B)). Alternating white and black diagonals with velocity amplitudes less than 1 km/s occur quasi-periodically in the space-time plot, representing recurring upward propagating Doppler velocity perturbations along the selected track (the plume structure). The inclination of the diagonal bands represents the phase speed of these line-of-sight perturbations along the track, as outlined by the blue dot-dashed lines (which again have been shifted just to the right of the target features) in Figure~\pref{f2} (B). Some of the perturbations appear to get weaker at higher heights. This is most likely due to a combination of increased noise levels in the data, reducing the signal-to-noise but also real, physical decay of the perturbations due to mode coupling of the propagating transverse waves to azimuthal Alfv\'en waves, as recently modelled by e.g.~\cite{Pascoe2010,Pascoe2013}.

Figure~\pref{f3} shows similar space-time plots as the one in Figure~\pref{f2} (B) but after subtracting 8 min, 14 min and 17 min smoothed versions, respectively, from the original, raw data. These plots clearly show that the quasi-periodic perturbations are real physical features and not an artefact of the chosen smoothing interval as the recurring 3-8 min Doppler velocity perturbations exhibit almost the same behavior even when the smoothing interval is changed from 5 to 17 min. 

To verify the phase speed of the propagating perturbations, we employ the same cross-correlation method  as\cite{Tomczyk_McIntosh_2009}, which fits the lead/lag times versus the distance along the selected track, relative to the midpoint of the track with a straight line. The propagation speed is estimated to be about 740$\pm$107 km/s (with a cross-correlation factor about 0.6), consistent with the \alfvenic{} wave speed obtained by \cite{Tomczyk_McIntosh_2009} and much larger than the typical sound speed \citep[$\sim$ 100 km/s, e.g.,][]{DeForest_Gurman_1998} in solar polar plumes. 

A spatially-averaged cut across the plume in the region indicated by the two horizontal dashed red lines in Figure~\pref{f2} (B) is shown in panel (C). Individual events with amplitudes around 0.5 km/s occur quasi-periodically with a recurrence rate from 3 to 8 minutes, consistent with our visual estimate from the space-time plots. Wavelet analysis \citep{Torrence_Compo_1998} on the time series reveals a dominant period of about $3.7 \pm 0.7$ minutes. As in previous studies, no corresponding intensity perturbations (with similar periodicity) are observed in the CoMP or AIA intensity observations, indicating the largely non-compressive (\alfvenic) nature of the disturbances.

Subsequently, we also perform the same analysis on the AIA observations. Figures~\ref{f4} (A) and (C) show the corresponding space-time plots of the AIA 193 \AA{} and 171 \AA{} intensity observations along the selected track, respectively.

The AIA 193 \AA{} space-time plot clearly shows recurring, propagating intensity perturbations (diagonal green and red bands), but with a much longer period than the CoMP Doppler velocity perturbations observed along the same track, suggesting that several different physical processes might be taking place along (or within) this plume structure. The upward propagating AIA intensity perturbations are of the order of 1\% (compared to the background intensity, taken as a 5-minute smoothed version of the initial data) and have propagation speeds of about 130 km/s (see \cite{Banerjee2011} for a review of similar intensity disturbances propagating along coronal plumes). Wavelet analysis reveals a dominant period of about 12 minutes, which is indeed considerably longer than that of the Doppler perturbations. The AIA 171 \AA{} passband data reveal almost identical results.
 
Applying the same approach on 13 other plume structures in the south polar coronal hole yields similar results, namely the combination of fast-propagating Doppler velocity perturbations and low amplitude, slower intensity perturbations travelling along all the plumes. Table~\ref{tb} shows the observed properties for all 14 plumes. Statistical analysis clearly shows the very different propagating speeds (834$\pm$203 km/s vs.~120$\pm$18 km/s) and periods (4.8$\pm$0.9 min vs.~15.2$\pm$5.0 min). Careful inspection of the movies provided in the online material suggests the ubiquitous presence of these high-speed Doppler velocity perturbations and weaker and slower intensity perturbations along most plumes in the field-of-view.

\begin{table}
\begin{center}
\linespread{1.5} \caption{Properties of the Perturbations Observed along 14 Plumes observed on December 30 2011.}

\begin{tabular}{c|ccc|ccc}
\hline
NO. & \multicolumn{3}{c|}{\comp{} Doppler Velocity} & \multicolumn{3}{c}{\aia{} 193\AA{} Intensity} \\
\hline
& $v_{d}\ (km/s)$  & $P_d\ (min)$ & $v_{pd}\ (km/s)$ & $I_{e}\ (\%)$ & $P_a \ (min)$ & $v_{pa}\ (km/s)$  \\
\hline
1  &  0.42  &  4.9  &  1021.7  &  0.97  &  20.1  &  113.6 \\
2  &  0.53  &  3.7  &  743.5  &  1.00  &  11.9  &  131.7 \\
3  &  0.58  &  5.6  &  662.9  &  1.13  &  14.5  &  102.8 \\
4  &  0.51  &  4.3  &  1149.0  &  0.92  &  15.5  &  80.7 \\
5  &  0.50  &  5.0  &  976.9  &  0.99  &  14.1  &  114.3 \\
6  &  0.58  &  5.6  &  607.5  &  0.51  &  14.4  &  116.5 \\
7  &  0.41  &  4.9  &  655.6  &  0.41  &  17.0  &  125.3 \\
8  &  0.47  &  4.3  &  654.8  &  0.54  &  14.1  &  114.7 \\
9  &  0.38  &  5.4  &  932.1  &  0.70  &  20.4  &  131.6 \\
10  &  0.45  &  6.9  &  818.0  &  1.27  &  14.0  &  100.8 \\
11  &  0.40  &  5.9  &  690.5  &  0.69  &  10.7  &  130.8 \\
12  &  0.39  &  5.8  &  633.5  &  0.65  &  15.0  &  147.6 \\
13  &  0.43  &  4.6  &  906.5  &  1.40  &  16.0  &  120.2 \\
14  &  0.43  &  4.4  &  1225.0  &  0.47  &  13.6  &  149.0 \\
\hline
\end{tabular}\\
\textbf{Notes.} $v_d$, $P_d$, $v_{pd}$ are the peak Doppler velocity difference, the period and the phase speed of the perpendicular wave motions observed in the CoMP Doppler velocity images, respectively. $I_e$, $P_{a}$ and $v_{pa}$ are the peak intensity enhancement, the periodicity and the propagation speed of the longitudinal intensity disturbances in the  AIA 193\AA{} images, respectively. Example no 2 is the plume discussed in detail in this paper.\label{tb}
\end{center}
\end{table}

\section{Discussions}\label{dis}

The observational analysis presented above demonstrated the co-existence of two characteristic types of perturbations in solar polar plumes: transverse (line-of-sight) Doppler velocity perturbations observed by the CoMP instrument, interpreted as propagating \alfvenic{} waves and weak, parallel (longitudinal), quasi-periodic, propagating disturbances in the AIA intensities.  The combination of these different types of perturbations, co-existing on the same (plume) structure indicates a richness of different physical processes taking place, at or near the potential source region of the fast solar wind. A similar co-existence of different perturbations was also found by \cite{Threlfall2013} along large coronal loops.

The observed \alfvenic{} motions propagate with phase speeds around $830 \pm 200$ km/s, similar to \alfvenic{} perturbations observed before in the solar corona \citep[e.g.,][]{Tomczyk_McIntosh_2009, DeMoortel2014}. The (Doppler) perturbation amplitudes are of the order of 0.5 km/s, with no obvious corresponding (observed) intensity variation. To first order, the energy flux carried by these waves can be estimated as $F_W=\rho<v^2>V_{phase}$, where $\rho$ is the plasma density, $v$ the velocity amplitude and $V_{phase}$ the phase speed (see also \cite{Goossens2013} for a discussion on energy content in \alfvenic{} waves). As demonstrated in \cite{McIntosh_DePontieu_2012} and \cite{IDM-Pascoe2012}, perturbation amplitudes obtained from the Doppler velocities could substantially underestimate the true velocity amplitudes (and hence energy flux) due to the relatively low (spatial) resolution and/or the line-of-sight superposition effects. Using observed line-widths and comparing with Monte-Carlo simulations, \cite{McIntosh_DePontieu_2012} estimated that the true amplitudes of the perturbations could be of the order of 25-56 km/s. With typical values in solar polar plumes for the electron number density of the order of $10^8\,$cm$^{-3}$, and the phase speed $\sim$830 km/s, we obtain an energy flux carried by the observed \alfvenic{} waves of about 80-400 W m$^{-2}$, sufficient to balance the estimated loss of about $\sim$100 W m$^{-2}$ \citep{Withbroe1977}.

Upward-propagating intensity perturbations are observed in the AIA 193 and 171 \AA{} images, co-existing on the plumes with the \alfvenic{} wave motions, but with much slower propagation speeds of $\sim$120 km/s, which is of the order of the theoretically expected slow magnetoacoustic speed in the solar polar region \citep[e.g.,][]{DeForest_Gurman_1998}. Similar low-amplitude, quasi-periodic intensity perturbations are commonly observed in the solar atmosphere and we refer the reader to, for example, \cite{IDM09} or \cite{Banerjee2011} for a review. Early observations (mostly using only imaging observations) interpreted the observed perturbations as propagating, slow magneto-acoustic waves. However, recent spectroscopic observations \citep[e.g.,][]{DePontieu2009, McIntosh2010} have revealed a more complex picture, indicating a mass motion component. Forward modelling of numerical simulations by \cite{IDM2015} highlighted the fundamental difficulty of distinguishing between these two different interpretations (slow propagating waves and quasi-periodic upflows). However, regardless of whether these intensity perturbations are  quasi-periodic upflows or slow magnetoacoustic waves, the intensity (density) enhancements might cause reflection of the fast \alfvenic{} waves and the interaction of the reflected wave trains could lead to the onset of an \alfvenic{} turbulent cascade. Such a turbulent cascade would enhance the dissipation of \alfvenic{} waves, potentially heating the local corona or accelerating the (fast) solar wind \citep[e.g.,][]{Velli1993, Matthaeus1999, Verdini2010, Oughton2001, Cranmer2005, vanBallegooijen2011}.

\section{Conclusions}\label{conc}

We studied combined CoMP and SDO/AIA observations of 14 plumes in the south polar coronal hole on December 30 2011. Detailed analysis of the characteristic properties of the (propagating) \alfvenic{} motions and quasi-periodic intensity perturbations highlight the physical richness of the processes taking place in polar plumes, at the base of the fast solar wind. The (perpendicular) \alfvenic{} waves have an average velocity amplitude of 0.5 km/s, projected phase speed (in the plane of the sky) of 830 km/s and periods of about 5 minutes. The (parallel) intensity perturbations observed along the {\em same} structures have an apparent (projected) speed of 120 km/s (in both the 171 \AA{} and 193 \AA{} passbands) and a much longer periodicity of about 15 minutes.
 
These observations potentially illustrate a scenario where the \alfvenic{} motions are propagating along, and through, an (longitudinally) inhomogeneous density structure (the polar plume) such that the combination could be a natural progenitor of the MHD turbulence required to accelerate the fast solar wind. However, further direct evidence for the existence of such a turbulent cascade within the plume structures is still needed and might be possible with higher resolution and cadence observations from, for example, IRIS. Theoretical modelling is required to investigate whether the small-amplitude intensity perturbations ($\sim 1$\%) are sufficiently effective in reflecting the \alfvenic{} waves to indeed establish a turbulent cascade.

\acknowledgements
Acknowledgements JL was a student visitor at HAO. JL acknowledges the financial support for his visit to HAO from the Chinese Scholarship Council (CSC). SWM appreciates the support of the Royal Society of Edinburgh and hospitality of staff in the University of St Andrews School of Mathematics and Statistics during his extended visit in the summer of 2014. NCAR is sponsored by the National Science Foundation. CoMP data can be found at the \url[http://www2.hao.ucar.edu/mlso/mlso-home-page]{MLSO Website}. We acknowledge support from NSFC 41131065, 41121003, 973 Key Project 2011CB811403 and CAS Key Research Program KZZD-EW-01-4. We also acknowledge support from NASA contracts NNX08BA99G, NNX11AN98G, NNM12AB40P, NNG09FA40C ({\em IRIS}), and NNM07AA01C (\hinode).  The research leading to these results has also received funding from the European Commission Seventh Framework Programme (FP7/ 2007-2013) under the grant agreement SOLSPANET (project No. 269299, \url{www.solspanet.eu/solspanet}).  

\begin{small}
\bibliographystyle{agufull}

\end{small}

\end{document}